\documentclass[12pt,thmsa]{article}
\usepackage{amssymb}

\begin{document}

\title{\textbf{Unorthodox properties of critical clusters}}
\author{A. Robledo \thanks{%
E-mail address: robledo@fisica.unam.mx } \\
Instituto de F\'{\i}sica,\\
Universidad Nacional Aut\'{o}noma de M\'{e}xico,\\
Apartado Postal 20-364, M\'{e}xico 01000 D.F., Mexico.}
\maketitle

\begin{abstract}
We look at the properties of clusters of order parameter $\phi $ at critical
points in thermal systems and consider their significance to
statistical-mechanical ground rules. These properties have been previously
obtained through the saddle-point approximation in a coarse-grained
partition function. We examine both static and dynamical aspects of a single
large cluster and indicate that these properties fall outside the canonical
Boltzmann-Gibbs (BG) scheme. Specifically: 1) The faster than exponential
growth with cluster size of the space-integrated $\phi $ suggests
nonextensivity of the BG entropy but extensivity of a $q$-entropy
expression. 2) The finding that the time evolution of $\phi $ is described
by the dynamics of an intermittent nonlinear map implies an atypical
sensitivity to initial conditions compatible with $q$-statistics and
displays an 'aging' scaling property. 3) Both, the approach to criticality
and the infinite-size cluster limit at criticality manifest through a
crossover from canonical to $q$-statistics and we discuss the nonuniform
convergence associated to these features.

Key words: critical clusters, saddle-point approximation, $q$-entropy,
intermittency, aging

PACS: 64.60.Ht, 75.10.Hk, 05.45.Ac, 05.10.Cc
\end{abstract}

\section{Introduction}

An unanticipated -- and remarkable -- relationship between intermittency and
critical phenomena has been recently suggested \cite{athens1} \cite{athens2}%
. This development brings together fields of research in nonlinear dynamics
and condensed matter physics, specifically, the dynamics in the proximity of
an incipiently chaotic attractor \cite{schuster1} appears associated to the
dynamics of fluctuations of an equilibrium state with well-known scaling
properties \cite{critical1}. Here we examine this connection in some detail
with special attention to several unorthodox properties, such as, the
extensivity of entropy of fractal clusters, the anomalous - faster than
exponential - sensitivity to initial conditions, and the aging scaling
features in time evolution.

Our interest is on the local fluctuations of a system undergoing a second
order phase transition, for example, in the Ising model as the magnetization
fluctuates and generates magnetic domains on all size scales at its critical
point. In particular, the object of study is a single cluster of order
parameter $\phi $ at criticality. This is described by a coarse-grained free
energy or effective action, like in the Landau-Ginzburg-Wilson (LGW)
continuous spin model portrayal of the equilibrium configurations of Ising
spins at the critical temperature and zero external field. As we shall see,
a cluster of radius $R$ is an unstable configuration whose amplitude grows
in time and eventually collapses when an instability is reached. This
process has been shown \cite{athens1} \cite{athens2} to be described by a
nonlinear map with tangency and feedback features \cite{schuster1}, such
that the time evolution of the cluster is given in the nonlinear system as a
laminar episode of intermittent dynamics.

The method employed to determine the cluster's order parameter profile $\phi
(\bold{r})$ assumes the saddle-point approximation of the
coarse-grained partition function $Z$, so that $\phi (\bold{r})$ is
its dominant configuration and is determined by solving the corresponding
Euler-Lagrange equation. The procedure is equivalent to the density
functional approach for stationary states in equilibrium nonuniform fluids.
The solution found for $\phi (\bold{r})$ is similar to an instanton in
field theories \cite{yangmills1}, and from its thermal average (evaluated by
integrating over its amplitude $\phi _{0}$, the remaining degree of freedom
after its size $R$ has been fixed) interesting properties have been derived.
These are the fractal dimension of the cluster \cite{athens3} \cite{athens4}
and the intermittent behavior in its time evolution \cite{athens1} \cite
{athens2}. Both types of properties are given in terms of the critical
isotherm exponent $\delta $.

As we describe below, the dominance of $\phi (\bold{r})$ in $Z$
depends on a condition that can be expressed as an inequality between two
lengths in space. This is $r_{0}\gg R$, where $r_{0}$ is the location of a
divergence in the expression for $\phi (\bold{r})$ that decreases as
an inverse power of the cluster amplitude $\phi _{0}$. When $r_{0}\gg R$ the
profile is almost horizontal but for $r_{0}\gtrapprox R$ the profile
increases from its center faster than an exponential. It is this feature
that gives the cluster some atypical properties that we present and discuss
here. These properties relate to the dependence of i) the number of cluster
configurations on size $R$, and ii) the sensitivity to initial conditions $%
\xi _{t}$ of order-parameter evolution on time $t$.

The above-mentioned properties appear to be at odds with the usual
Boltzmann-Gibbs (BG) statistics but suggest compatibility \cite{robledo1}
with one of its generalizations known as $q$-statistics \cite{tsallis0} \cite
{tsallis1}. A condition for these properties to arise is criticality but
also is the circumstance that phase space has only been partially
represented by selecting only dominant configurations. Hence, the motivation
to examine this problem rests on explaining the physical and methodological
basis under which proposed generalizations of the BG statistics may apply.
For this reason, we begin in Section 2 by recalling the basic framework for
the Landau approach, and, in Section 3, we review details of the derivation
of the dominant configuration. Then, in Section 4 we consider an estimate
for the numbers of configurations that give rise to a cluster and check on
the extensivity of entropy expressions. In section 5 we review the
connection between the cluster time evolution and the nonlinear dynamics of
intermittency, so that in Section 6 we can discuss again the manifestation
of $q$-statistics. In Section 7 we summarize and discuss our results.

\section{A Landau approach for single clusters}

We start our account by recalling that the Landau approach is backed by a
two-stage calculation strategy for the partition function $Z$ of the system.
In the first, a sum is made over the microscopic configurations that lead to
a specific form for the order parameter $\phi (\bold{r})$ (where $%
\bold{r}$ is the spatial position) to obtain a partial, $\phi (%
\bold{r})$-dependent, result $Z_{\phi }$. The second stage consists of
summing up $Z_{\phi }$ over all possible forms $\phi (\bold{r})$, so
that 
\begin{equation}
Z=\int D[\phi ]Z_{\phi },  \label{part1}
\end{equation}
where a path integration over the configurations for $\phi $ is indicated.
The partial partition function $Z_{\phi }$ is then written as 
\begin{equation}
Z_{\phi }=\Omega \lbrack \phi ]\exp (-E[\phi ]/kT),  \label{part2}
\end{equation}
where $E[\phi ]$ is the energy of the system when the order parameter takes
the fixed form $\phi (\bold{r})$, and $\Omega \lbrack \phi ]$ is the
number of microscopic configurations for that $\phi (\bold{r})$.
Taking logarithms to both sides of Eq. (\ref{part2}) gives 
\[
\Psi \lbrack \phi ]=\frac{E[\phi ]}{T}-S[\phi ], 
\]
where $\Psi \lbrack \phi ]$ is the so-called LGW free energy and $S[\phi
]=\ln \Omega \lbrack \phi ]$ is an 'entropy' term associated to $\Omega
\lbrack \phi ]$ (Boltzmann's constant has been taken as unity). In practice
the evaluation of 
\begin{equation}
Z=\int D[\phi ]\exp (-\Psi \lbrack \phi ])  \label{part3}
\end{equation}
is considered with $\Psi \lbrack \phi ]$ written as a spatial integral over
some function of $\phi (\bold{r})$ that involves a square gradient
(perhaps higher-order derivatives) and powers of $\phi (\bold{r})$.

At criticality the LGW free energy takes the form 
\begin{equation}
\Psi _{c}[\phi ]=a\int dr^{d}\left[ \frac{1}{2}\left( \nabla \phi \right)
^{2}+b\left\vert \phi \right\vert ^{\delta +1}\right] ,  \label{lgw1}
\end{equation}
$\delta $ is the critical isotherm exponent and $d$ is the spatial dimension
($\delta =5$ for the $d=3$ Ising model with short range interactions). The
coefficients $a$ and $b$ are dimensionless couplings and $\phi =\lambda ^{-d}%
\widetilde{\phi }$ and $r=\lambda \widetilde{r}$ are dimensionless
quantities, $\lambda $ is the ultraviolet cutoff that fixes the coarse
graining scale, $\widetilde{\phi }$ is the order parameter (e.g.
magnetization per unit volume) and $\widetilde{r}$ represents a spatial
position coordinate with dimension of length. Out of criticality $\Psi
\lbrack \phi ]$ takes the form $\Psi \lbrack \phi ]=\int dr^{d}\left[ 1/2\
\left( \nabla \phi \right) ^{2}+1/2\ r_{0}\phi ^{2}+u_{0}\phi ^{4}\right] $,
where $r_{0}=a_{0}t$, \ $a_{0}$ and $u_{0}>0$ are constants and $t$ is the
reduced temperature distance to the critical point. The ordering field has
been set to zero \cite{chaikin1}.

To study the coarse-grained configurations of a single cluster, a subsystem
of finite size $R$ is considered in the neighborhood of $\bold{r}=%
\bold{0}$ and there are no restrictions imposed to the value of $\phi $
at its boundaries with the rest of the (infinite-sized) physical system. In
this case the integration in Eq. (\ref{lgw1}) is performed over the
subsystem's volume $V$.

\section{Cluster properties from dominant configurations}

We consider a one dimensional system with unspecified range of interactions,
as in this case the procedure is more transparent and the expressions for
the relevant quantities more easily derived. The analysis can be carried out
for higher dimensions with no further significant assumptions and with
comparable results \cite{athens1}-\cite{athens4}. The LGW free energy reads
now 
\begin{equation}
\Psi _{c}[\phi ]=a\int_{0}^{R}dx\left[ \frac{1}{2}\left( \frac{d\phi }{dx}%
\right) ^{2}+b\left\vert \phi \right\vert ^{\delta +1}\right] ,  \label{lgw2}
\end{equation}
and we adopt the saddle-point approximation - valid for $a\gg 1$ - to
circumvent the nontrivial task of carrying out the path integration in Eq. (%
\ref{part1}). The saddle-point configurations are obtained from the
Euler-Lagrange equation 
\begin{equation}
\frac{d^{2}\phi }{dx^{2}}=-\frac{dV}{d\phi },  \label{EL1}
\end{equation}
where $V=-b\left\vert \phi \right\vert ^{\delta +1}$. It is helpful to
recall the classical motion analog of this problem, a particle in time $x$
under a potential $V$. Integration of Eq. (\ref{EL1}) yields 
\begin{equation}
U=\frac{1}{2}\left( \frac{d\phi }{dx}\right) ^{2}-b\left\vert \phi
\right\vert ^{\delta +1},  \label{energy1}
\end{equation}
where the constant $U$ is the total particle's energy. Subsequent
integration of Eq. (\ref{energy1}) with $U=0$ leads to profiles for critical
clusters of the form \cite{athens1}-\cite{athens4} 
\begin{equation}
\phi (x)=A\left\vert x-x_{0}\right\vert ^{-2/(\delta -1)},  \label{profile1}
\end{equation}
where 
\[
A=\left[ \sqrt{b/2}\left( \delta -1\right) \right] ^{-2/(\delta -1)} 
\]
and 
\begin{equation}
x_{0}=\left[ \sqrt{b/2}\left( \delta -1\right) \right] ^{-1}\phi
_{0}^{-(\delta -1)/2},  \label{sing1}
\end{equation}
where $x_{0}$ is a system-dependent reference position and $\phi _{0}=$ $%
\phi (0)$. The value of $\phi $ at the edge of the cluster is $\phi _{R}=$ $%
\phi (R)$ and the cluster free energy is 
\begin{equation}
\Psi _{c}[\phi ]=2ab\int_{0}^{R}dx\ \phi (x)^{\delta +1}.  \label{lgw3}
\end{equation}
This family of solutions give the largest contributions to $Z$ and are the
analogs of instantons in semiclassical quantum theories, i.e. configurations
that minimize the Yang-Mills action \cite{yangmills1}.

Similar solutions are obtained for small $U\approx 0$ \cite{athens1} \cite
{athens2} where now the position of the singularity $x_{0}$ depends also on $%
U$. These solutions enter $Z$ with a weight $\exp (-\alpha R\left| U\right|
) $ and therefore their relevance diminishes as $\left| U\right| $ increases.

\section{Extensivity of critical cluster entropy}

We comment now on the nature of the profile $\phi (x)$ in Eq. (\ref{profile1}%
). This function, as well as those solutions for $U\simeq 0$, can be
rewritten in the form 
\begin{equation}
\phi (x)=\phi _{0}\exp _{q}(kx),  \label{q-exp1}
\end{equation}
where $\exp _{q}(x)$ is the $q$-exponential function $\exp _{q}(x)\equiv
\lbrack 1-(q-1)x]^{-1/(q-1)}$ with $q=$ $(1+\delta )/2$ and $k=\sqrt{2b}\phi
_{0}^{(\delta -1)/2}$. Because $\delta >1$ one has $q>1$ and $\phi (x)$
grows faster than an exponential as $x\rightarrow x_{0}$ and diverges at $%
x_{0}$. It is important to notice that only configurations with $R\ll x_{0}$
have a nonvanishing contribution to the path integration in Eq. (\ref{part1}%
) \cite{athens1} \cite{athens2} and that these configurations vanish for the
infinite cluster size system. There are some characteristics of nonuniform
convergence in relation to the limits $R\rightarrow \infty $ and $%
x_{0}\rightarrow \infty $, a feature that is significant for our connection
with $q$-statistics. By taking $\delta =1$ the system is set out of
criticality, then $q=1$ and the profile $\phi (x)$ becomes the exponential $%
\phi (x)=\phi _{0}\exp (k_{0}\ x)$, $k_{0}=\sqrt{a_{0}t}$. (Note that in
this case $x_{0}\rightarrow 0$).

The quantity
\begin{equation}
\Phi (R)=\int_{0}^{R}dx\phi (x),  \label{totalmag1}
\end{equation}
or total 'magnetization' of the cluster, is given by
\begin{equation}
\Phi (R)=\Phi _{0}\left\{ \left[ \exp _{q}(kR)\right] ^{2-q}-1\right\} ,\
R<x_{0},  \label{totalmag2}
\end{equation}
where $\Phi _{0}=$sgn$(3-\delta )[2\phi _{0}/(\delta -3)k]$ with $%
q=(1+\delta )/2$. (We have not shown the special case $\delta =3$). For $%
\delta =1$, one has $\Phi (R)=\phi _{0}k_{0}^{-1}[\exp (k_{0}R)-1]$. The
rate at which $\Phi (R)$ grows with $R$, $d\Phi (R)/dR$, is necessarily
equal to $\phi _{R}$, the value of $\phi (x)$ at the edge of the cluster,
therefore 
\begin{equation}
\frac{d\Phi (R)}{dR}=\phi _{0}\exp _{q}(kR),\ R<x_{0},  \label{rate1}
\end{equation}
while for $\delta =1$ it is $d\Phi (R)/dR=\phi _{0}\exp (k_{0}R)$.

The expressions above may be used to estimate the dependence on cluster size 
$R$ of the number of microscopic configurations $\Omega \lbrack \phi ]$ that
make up the partial partition function $Z_{\phi }$ in Eq. (\ref{part2}) for
the dominant coarse-grained $\phi (x)$. This dependence may be obtained in a
way analogous to that of how is obtained the dependence with time of the
number of configurations $\Omega $ for an ensemble of trajectories in a
one-dimensional dynamical system. Here 'trajectory positions' are given by
the values of $\phi $ in microscopic configurations and `time' is given by
the cluster size $R$. Initially adjacent positions stay adjacent and $\Omega 
$ is almost constant but at later times they spread and $\Omega $ increases
rapidly. For chaotic orbits the increment is exponential \cite{schuster1}
but for marginally chaotic orbits at the tangent bifurcation $\Omega $
increases as a $q$-exponential with $q>1$ \cite{robledo2} \cite{baldovin1}.
The ensemble of trajectories is initially contained in the interval $[0,\phi
_{0}]$ and at time $R$ they occupy the interval $[0,\phi _{R}]$, therefore
we assume 
\begin{equation}
\Omega (R)\sim \phi _{0}^{-1}d\Phi (R)/dR=\phi _{0}^{-1}\phi _{R}.
\label{conf1}
\end{equation}
The results in the following Sections provide a justification for this
choice.

Then, it is significant to note that the Tsallis entropy \cite{tsallis0}, 
\begin{equation}
S_{q}=\ln _{q}\Omega \equiv \frac{\Omega ^{1-q}-1}{1-q},  \label{tsallis1}
\end{equation}
(where $\ln _{q}y\equiv (y^{1-q}-1)/(1-q)$ is the inverse of $\exp _{q}(y)$%
), when evaluated for $\Omega \sim \exp _{q}(kR)$ complies with the
extensivity property $S_{q}\sim R$ \cite{tsallis2}, while the BG entropy 
\begin{equation}
S_{BG}(t)=\ln \Omega ,  \label{bg1}
\end{equation}
obtained from $S_{q}$ when $q=1$, when evaluated for $\Omega \sim \exp
(k_{0}R)$ complies also with the extensivity property $S_{1}\sim R$.

\section{Cluster instability and intermittency}

The profile $\phi (x)$ given by Eqs. (\ref{profile1}) or (\ref{q-exp1})
describes a fluctuation of the critical equilibrium state of the infinite
system with average $\left\langle \overline{\phi }\right\rangle =0$. In a
coarse-grained time scale the cluster is expected to evolve by increasing
its amplitude $\phi _{0}$ and size $R$ because the subsystem studied
represents an environment with unevenness in the states of the microscopic
degrees of freedom (e.g. more spins up than down). Increments in $\phi _{0}$
for fixed $R$ takes the position $x_{0}$ for the singularity closer to $R$
and the almost constant shape $\phi (x)\simeq \phi _{0}$ for $x_{0}\gg R$ is
eventually replaced by a faster than exponential shape $\phi (x)$, while, as
mentioned, the dominance of this configuration in $Z$ decreases accordingly
and rapidly so. When the divergence is reached at $x_{0}=$ $R$ the profile $%
\phi (x)$ no longer describes the spatial region where the subsystem is
located. But a subsequent fluctuation would again be represented by a
cluster $\phi (x)$ of the same type. From this renewal process we obtain a
picture of intermittency. A similar situation would occur if $R$ is
increased for fixed $\phi _{0}$.

Indeed, a link was revealed \cite{athens1} \cite{athens2} between the
fluctuation properties of a critical cluster described by Eqs. (\ref
{profile1}) and (\ref{q-exp1}) and the dynamics of marginally chaotic
intermittent maps. This connection was demonstrated in different ways in
Refs. \cite{athens1} \cite{athens2} by considering the properties of the
thermal average
\begin{equation}
\left\langle \Phi (R)\right\rangle =Z^{-1}\int D[\phi ]\ \Phi (R)\ \exp
(-\Psi _{c}[\phi ]),  \label{avPhi1}
\end{equation}
for fixed $R$. When $x_{0}\gg R$ the profile is basically flat $\phi
(x)\simeq \phi _{0}$, the LGW free energy is $\Psi _{c}\simeq 2abR\phi
_{0}^{\delta +1}$, and the path integral in Eq. (\ref{avPhi1}) becomes an
ordinary integral over $0\leq \phi \leq \phi _{0}$. One obtains 
\begin{equation}
\left\langle \Phi (R)\right\rangle \simeq \frac{\phi _{0}R}{2}\exp \left(
-uR\phi _{0}^{\delta +1}\right) ,  \label{avPhi2}
\end{equation}
where $u=2ab(\delta +1)/(\delta +2)(\delta +3)$.

The procedure that resembles the picture given in the beginning of this
Section is to consider the value of $\left\langle \Phi \right\rangle $ at
successive times $t=0,1,...$, and assume that this quantity changes by a
fixed amount $\mu $ per unit time, that is
\begin{equation}
\left\langle \Phi _{t+1}\right\rangle =\left\langle \Phi _{t}\right\rangle
+\mu .  \label{map1}
\end{equation}
Making use of Eq. (\ref{avPhi2}) one obtains \cite{athens3} \cite{athens4}
for small values of $\phi _{t}$ the map
\begin{equation}
\phi _{t+1}=\epsilon +\phi _{t}+\nu \phi _{t}^{\delta +1},  \label{tangent1}
\end{equation}
where the shift parameter is $\epsilon \sim R^{-1}$ and the amplitude of the
nonlinear term is $\nu =u\mu $.

Eq. (\ref{tangent1}) can be recognized as that describing the intermittency
route to chaos in the vicinity of a tangent bifurcation \cite{schuster1}.
The complete form of the map \cite{athens1} \cite{athens2} displays a
'superexponentially' decreasing region that takes back the iterate close to
the origin in approximately one step. Thus the parameters of the thermal
system determine the dynamics of the map. The mean number of iterations in
the laminar region was seen to be related to the mean magnetization within a
critical cluster of radius $R$. There is a corresponding power law
dependence of the duration of the laminar region on the shift parameter $%
\epsilon $ of the map \cite{athens1}. For $\epsilon >0$ the (small) Lyapunov
exponent is simply related to the critical exponent $\delta $ \cite{athens2}.

\section{ Intermittency and \textit{q}-statistics}

At the tangent bifurcation, the intermittency route to chaos, the ordinary
Lyapunov exponent $\lambda _{1}$ vanishes and the sensitivity to initial
conditions $\xi _{t}\equiv \left\vert dx_{t}/dx_{in}\right\vert $ (where $%
x_{t}$ is the orbit position at time $t$ given the initial position $x_{in}$
at time $t=0$) is no longer given by the BG law $\xi _{t}=\exp (\lambda
_{1}t)$ but acquires either a power or a super-exponential law \cite
{robledo2} \cite{baldovin1}. In this case $\xi _{t}$ is given by the $q$%
-exponential expression, 
\begin{equation}
\xi _{t}=\exp _{Q}(\lambda _{Q}t)\equiv \lbrack 1-(Q-1)\lambda
_{Q}t]^{-1/(Q-1)},  \label{qsensitivity}
\end{equation}
containing the entropic index $Q$ and the $q$-generalized Lyapunov
coefficient $\lambda _{Q}$. In the limit $Q\rightarrow 1$ Eq. (\ref
{qsensitivity}) reduces to the ordinary BG law. See \cite{robledo2} and \cite
{baldovin1} for a more rigorous description of the theory and related issues.

Assisted by the known renormalization group (RG) treatment for the tangent
bifurcation \cite{schuster1}, the formula for $\xi _{t}$ has been rigorously
derived \cite{robledo2} \cite{baldovin1} and found to comply with Eq. (\ref
{qsensitivity}). The tangent bifurcation is usually studied by means of the
map 
\begin{equation}
f(x)=\epsilon +x+\nu \left\vert x\right\vert ^{z}+o(\left\vert x\right\vert
^{z}),\ \nu >0,  \label{n-thf1}
\end{equation}
with nonlinearity $z>1$ in the limit $\epsilon \rightarrow 0$. The
associated RG fixed-point map $x^{\prime }=f^{\ast }(x)$ was found to be 
\begin{equation}
x^{\prime }=x\exp _{z}(\nu x^{z-1})=x[1-(z-1)\nu x^{z-1}]^{-1/(z-1)},\
\epsilon =0,  \label{fixed1}
\end{equation}
as it satisfies $f^{\ast }(f^{\ast }(x))=\alpha ^{-1}f^{\ast }(\alpha x)$
with $\alpha =2^{1/(z-1)}$ and has a power-series expansion in $x$ that
coincides with Eq. (\ref{n-thf1}) in the two lowest-order terms. (Above $%
x^{z-1}\equiv \left\vert x\right\vert ^{z-1}\mathrm{sgn}(x)$). The long time
dynamics is readily derived from the static solution Eq. (\ref{fixed1}), one
obtains 
\begin{equation}
\xi _{t}(x_{in})=[1-(z-1)\nu x_{in}^{z-1}t]^{-z/(z-1)},  \label{sensitivity2}
\end{equation}
and so, $Q=2-z^{-1}$ and $\lambda _{Q}(x_{in})=z\nu x_{in}^{z-1}$ \cite
{robledo2} \cite{baldovin1}. When $Q>1$ the left-hand side ($x<0$) of the
tangent bifurcation map, Eq. (\ref{n-thf1}), exhibits a weak insensitivity
to initial conditions, i.e. power-law convergence of orbits. However at the
right-hand side ($x>0$) of the bifurcation the argument of the $q$%
-exponential becomes positive and this results in a `super-strong'
sensitivity to initial conditions, i.e. a sensitivity that grows faster than
exponential \cite{baldovin1}. Comparison of Eq. (\ref{n-thf1}) with Eq. (\ref
{tangent1}) indicates the simple relation $z=\delta +1$.

There is an interesting scaling property displayed by $\xi _{t}$ in Eq. (\ref
{sensitivity2}) similar to the scaling property known as \emph{aging} in
systems close to glass formation. This property is observed in two-time
functions (e.g. time correlations) for which there is no time translation
invariance but scaling is observed in terms of a time ratio variable $%
t/t_{w} $ where $t_{w}$ is a 'waiting time' assigned to the time interval
for preparation or hold of the system before time evolution is observed
through time $t$. This property can be seen immediately in $\xi _{t}$ if one
assigns a waiting time $t_{w}$ to the initial position $x_{in}$ as $t_{w}=$ $%
x_{in}^{1-z}$. Eq. (\ref{sensitivity2}) reads now
\begin{equation}
\xi _{t}(t_{w})=[1-(z-1)\nu t/t_{w}]^{-z/(z-1)}.  \label{sensitivity3}
\end{equation}
The sensitivity $\xi _{t}$ for this critical attractor is dependent on the
initial position $x_{in}$ or, equivalently, on its waiting time $t_{w}$, the
closer $x_{in}$ is to the point of tangency the longer $t_{w}$ but the
sensitivity of all trajectories fall on the same $q$-exponential curve when
plotted against $t/t_{w}$. Aging has also been observed for the properties
of the same map but in a different context \cite{barkai1}.

\section{\textbf{Summary and discussion}}

We have examined the study of clusters at criticality in thermal systems by
means of the saddle-point approximation in the LGW free energy model \cite
{athens1} \cite{athens2} \cite{athens3} \cite{athens4}. The retention of
only one coarse-grained configuration leads to cluster properties that are
physically reasonable but also appear to fall outside the limits of validity
of the canonical theory. The fractal geometry and the intermittent behavior
of critical clusters obtained from this method \cite{athens1} \cite{athens2} 
\cite{athens3} \cite{athens4} are both consistent with equivalent properties
found for clusters at the critical points of the $d=2$ Ising and Potts
models \cite{2dIsing1} \cite{2dIsing2}. On the other hand, we found that the
entropy expression that provides the property of extensivity for our
estimate of the number of cluster configurations is not the usual BG
expression but that of $q$-statistics. Likewise, the nonlinear map and its
corresponding sensitivity to initial conditions linked to the intermittency
of clusters do not follow the fully-chaotic trajectories of BG statistics
but display the features of $q$-statistics.

With regards to the extensivity of entropy, what our assumptions and results
mean basically is that extensivity of entropy (BG or $q$-generalized) of a
cluster and the Kolmogorov-Sinai \cite{schuster1} linear growth with time of
entropy (BG or $q$-generalized) of trajectories for the attractor of a
nonlinear map are equivalent. The crossover from the $S_{q}$ to the $S_{BG}$
expressions is obtained when the system is taken out of criticality, because 
$\delta \rightarrow 1$ makes $q\rightarrow 1$. When at criticality the
crossover between $q$-statistics and BG statistics in the dynamical behavior
is related to the subsystem's size in the following way. We keep in mind
that the map shift parameter depends on the domain size as $\epsilon \sim
R^{-1}$. So, the time evolution of $\phi $ displays laminar episodes of
duration $<t>\sim \epsilon ^{-\delta /(\delta +1)}$ and the Lyapunov
coefficient in this regime is $\lambda _{1}\sim \epsilon $ \cite{athens2}.
Within the first laminar episode the dynamical evolution of $\phi (x)$ obeys 
$q$-statistics, but for very large times the occurrence of many different
laminar episodes leads to an increasingly chaotic orbit consistent with the
small $\lambda _{1}>0$ and BG statistics is recovered. As $R$ increases ($%
R\ll x_{0}$ always) the time duration of the $q$-statistical regime
increases and in the limit $R\rightarrow \infty $ there is only one
infinitely long laminar $q$-statistical episode with $\lambda _{1}=0$ and
with no crossover to BG statistics. On the other hand when $R>x_{0}$ the
clusters $\phi (x)$ are no longer dominant, for the infinite subsystem $%
R\rightarrow \infty $ their contribution to $Z$ vanishes and no departure
from BG statistics is expected to occur \cite{robledo1}.

In view of the results presented here the departure from BG statistics and
the applicability of $q$-statistics is due in part to the presence of the
long-ranged correlations in space and in time that take place at
criticality. These correlations give the integrand in the LGW $\Psi _{c}$ a
power-law dependence of the form $\left\vert \phi \right\vert ^{\delta +1}$
with $\delta >1$ (commonly $\delta \geq 3$ as $\delta =3$ gives the Gaussian
critical point) and this in turn determines the $q$-exponential expression
for $\phi (x)$ and the properties derived from it. On the other hand, the
neglect of all coarse-grained configurations other than the most dominant
implies that phase space has not been properly sampled, and that the ergodic
and mixing properties characteristic of equilibrium BG statistics are not
guaranteed. In this respect it is perhaps significant to notice that the
most important instance for which $q$-statistics has been to date rigorously
proved to hold is the so-called onset of chaos in unimodal nonlinear maps 
\cite{baldovin2} \cite{mayoral1}. This marginally chaotic attractor is
nonergodic and nonmixing and its dynamical properties exhibit
infinitely-ranged time correlations. The crossover to BG statistics in this
model can be induced via perturbation with noise. Recently \cite{robledo3},
the dynamical properties of the noise-perturbed onset of chaos have been
proved to be analogous to those of supercooled liquids close to
vitrification.

\textbf{Acknowledgments.} It is with much pleasure and appreciation that I
dedicate this work to Benjamin Widom. Partial support by DGAPA-UNAM and
CONACyT (Mexican Agencies) is acknowledged.

\bigskip

\end{document}